\documentclass[prl,showpacs,twocolumn]{revtex4}

\newcommand {\ryx}{\rho_{yx}}

\newcommand {\tmo}{Tb$_2$Mo$_2$O$_7$}
\newcommand {\hmo}{Ho$_2$Mo$_2$O$_7$}

\newcommand {\nmo}{Nd$_2$Mo$_2$O$_7$}
\newcommand {\tcmo}{(Tb$_{1-x}$Ca$_{x}$)$_2$Mo$_2$O$_7$}
\newcommand {\tcdmo}{(Tb$_{1-x}$Cd$_{x}$)$_2$Mo$_2$O$_7$}
\newcommand {\rmo}{$R_2$Mo$_2$O$_7$}
\usepackage[pdftex]{graphicx} % usually error, but arXiv ok
\usepackage{bm}
\usepackage{pifont}
\usepackage{amsmath,amssymb}
\usepackage{ulem}

\begin{document}

\title{Highly anisotropic geometrical Hall effect via {\it f-d} exchange fields \\ in doped pyrochlore molybdates}

\author{Hikaru Fukuda$^1$}
\author{Kentaro Ueda$^{1,\ast}$}
\author{Yoshio Kaneko$^{2}$}
\author{Ryosuke Kurihara$^{2,3}$}
\author{Atsushi Miyake$^{4}$}
\author{Kosuke Karube$^{2}$}
\author{Masashi Tokunaga$^{2,4}$}
\author{Yasujiro Taguchi$^{2}$}
\author{Yoshinori Tokura$^{1,2,5}$}

\affiliation{
$^1$ Department of Applied Physics, University of Tokyo, Bunkyo, Tokyo 113-8656, Japan\\
$^2$ RIKEN Center for Emergent Matter Science, Wako, Saitama 351-0198, Japan\\
$^3$ Department of Physics, Tokyo University of Science, Noda, Chiba 278-0022, Japan\\
$^4$ Institute for Solid State Physics (ISSP), University of Tokyo, Kashiwa, Chiba 277-8581, Japan\\
$^5$ Tokyo College, The University of Tokyo, Tokyo 113-8656, Japan}
%
%
%$^1$ Department of Applied Physics, University of Tokyo, Hongo 7-3-1, Bunkyo-ku, Tokyo 113-8656, Japan\\
%$^2$ RIKEN Center for Emergent Matter Science (CEMS), Wako 351-0198, Japan\\
%$^3$ Department of Physics, Tokyo University of Science, Noda, Chiba 278-0022, Japan\\
%$^4$ Institute for Solid State Physics (ISSP), University of Tokyo, Kashiwa, Chiba 2778581, Japan\\
%$^5$ Tokyo College, The University of Tokyo, Hongo 7-3-1, Bunkyo-ku, Tokyo 113-8656, Japan}
%\item[*] Electronic address: ueda@ap.t.u-tokyo.ac.jp
\date{May 11, 2022}

\begin{abstract}
When a conduction electron couples with a non-coplanar localized magnetic moment, the real-space Berry curvature is exerted to cause the geometrical Hall effect, which is not simply proportional to the magnetization.
So far, it has been identified in the case mostly where the non-coplanar magnetic order is present on the sublattice of conduction electrons.
%Here we show that the geometrical Hall effect is induced by non-coplanar exchange fields from the localized magnetic moments, i.e., in the Kondo coupling case without long-range magnetic order on the conduction electron sublattice. In hole-doped pyrochlore \tmo, the barely metallic state with Mo-sublattice conduction electrons shows the largely anisotropic geometrical Hall effect with respect to the applied magnetic field direction in accord with the field-induced magnitude and sign change of the real-space scalar spin chirality formed by the Tb local moments.
Here, we demonstrate that the geometrical Hall effect shows up even without long-range magnetic order of conduction electrons, as induced by non-coplanar exchange fields from the localized magnetic moments, in hole-doped phyrochlore molybdates.
We find that the geometrical Hall effect is markedly anisotropic with respect to the applied magnetic field direction, which is in good accordance with the field-dependent magnitude and sign change of the real-space scalar spin chirality of local Tb moments.
These results may facilitate the understanding of emergent electromagnetic responses induced by the Kondo-like coupling between conduction electrons and local spins in a broad material class.
\end{abstract}
%\pacs{71.30.+h, 72.80.Ga, 78.20.-e, 78.30.-j}
\newpage\newpage\newpage
\maketitle
%o
Non-collinear or non-coplanar complex magnetic structures in solids are proven to play a fundamentally important role in their spin-related quantum transport and multiferroic properties \cite{Tokura2014,Nagaosa2013}.
A representative example is the scalar spin chirality (SSC), which is defined by $\chi_{ijk} = \bm{S}_{i} \cdot (\bm{S}_{j}\times \bm{S}_{k})$ for three neighboring-site spins $\bm{S}_{i}, \bm{S}_{j}$, and $\bm{S}_{k}$.
As a conduction electron moves over a non-coplanar spin texture with finite SSC, it is endowed with the non-trivial quantum phase (Berry phase) and hence experiences the emergent magnetic field, which can be far beyond a real magnetic field \cite{Ye1999,Taguchi2001,Nagaosa2010}.
%The finite SSC endows the electron wave function with the non-trivial quantum phase (Berry phase), giving rise to the unconventional Hall effect termed geometrical Hall effect (GHE).
%Unconventional Hall effect, termed geometrical Hall effect or topological Hall effect, which is not simply proportional to the magnetization as opposed to the conventional anomalous Hall, is one of the most common outcome of SSC, as observed in many non-coplanar magnets \cite{Taguchi2001}.
One of the most common outcomes of SSC is the unconventional Hall effect, termed geometrical Hall effect (GHE)  \cite{Nagaosa2010}.
In the case where the mean free path is sufficiently longer than the magnetic period, both anomalous Hall effect (AHE) and GHE are appropriately captured by the momentum space picture where the anti-crossing points (e.g. Weyl nodes) are associated with the Berry curvature due to the spin-orbit coupling (Karplus-Luttinger intrinsic mechanism) and the SSC \cite{Ohgushi1999, 2003ScienceFang}.
In contrast, as the magnetic period is longer than the mean free path, electrons hop around the spin triad in the real space and feel the emergent field $\bm{B}_{\mathrm{eff}} = h\bm{n}_{ijk}\chi_{ijk}/eS$, where $\bm{n}_{ijk}$ is the normal vector to the spin triad plane and $S$ is the area of spin triad, as exemplified by the magnetic skyrmion lattice \cite{2009PRLNeubauer,Kanazawai2011}.
%A representative example is the skyrmion lattice observed in chiral-lattice magnets where the size of skyrmion with quantized SSC is usually large, e.g. 10nm-100nm \cite{Ye1999,Kanazawai2011}.
%Most recently, high skyrmion density induces a giant GHE in Gd$_2$PdSi$_3$, stimulating a considerable interest in frustrated non-coplanar magnets \cite{2019ScienceKurumaji}.
\\
\ \ \ Among a variety of non-coplanar magnets, pyrochlore molybdates \rmo\ ($R$ being a trivalent rare-earth or Y ion) offer an ideal platform to study the correlation between charge transport and non-coplanar magnetism, because of a variety of magnetic/electronic phases \cite{2000PRLKatsufuji,2009PRLIguchi,2010RMPGardner}. 
The pyrochlore lattice consists of corner-linked tetrahedra with $R$ ions and Mo ions, each of which is displaced by half a unit cell (Fig. 1(a)). 
%Owing to the local trigonal distortion of MoO${}_6$ octahedron, Mo $t{}_{2\mathrm{g}}$ orbitals are split into doublet $e^\mathrm{{\prime}_{g}}$ and singlet $a_{1\mathrm{g}}$ manifolds.
%Two Mo-$4d$ electrons are accommodated in the $t{}_{2\mathrm{g}}$ state; one in $e_{g}^{\prime}$ may be itinerant while the other in a${}_{1\mathrm{g}}$ state is localized \cite{2003PRBSolovyev}.
The most well investigated is \nmo\ with the possible largest-size $R$ ion and hence with relatively small electron correlation.
It is metallic in the whole temperature range and undergoes the ferromagnetic transition at 90 K, presumably due to the double-exchange-like mechanism of Mo-$4d$ electrons \cite{1989SSCAli}.
Additionally, the Nd-$4f$ magnetic moments, which host Ising anisotropy along the local [111] direction, begin to freeze below around 40 K. Therefore, the Mo spins are slightly tilted ($\sim$ 5-10${}^{\circ}$) via the $f$-$d$ magnetic interaction at low temperatures, resulting in the Mo spins with finite SSC (Fig. 1(b)) \cite{Taguchi2001}.
%Since the Nd-$4f$ magnetic moments are Ising anisotropy along the local [111] direction, the Mo spins are slightly tilted ($\sim$ 5-10${}^{\circ}$) via the $f$-$d$ magnetic interaction, resulting in the Mo spins with finite SSC (Fig. 1(b)).%他の文献も
Because of such anisotropy for $R$-$4f$ moments, various configurations are realized under the external magnetic field \cite{1997PRLHarris,2001ScienceBramwell}.
%One important feature for pyrochlores with the Ising type $4f$ moments is that the magnetic configuration depends on the field direction.
For instance, as the field is applied along [100] direction, two of four magnetic moments point out of the tetrahedron, while the other two inwards, termed the 2-in 2-out configuration.
On the other hand, the field along [111] favors the 3-in 1-out state, as displayed in the bottom panel in Fig. 1(b).
The anisotropy of Hall response in \nmo\ is understood in terms of the 2-in 2-out or 3-in 1-out like habit of the Mo-$4d$ spins transmitted from the corresponding change of Nd-$4f$ configuration \cite{Taguchi2003}.
Such a case where the Mo-$4d$ ferromagnetic orders are slightly modulated by the $R$-$4f$ moments is termed the strong coupling, which has been extensively studied so far \cite{Taguchi2001,2021PRBHirschberger}.
On the other hand, little is known for the weak coupling regime, where the Mo conduction electrons exhibit no spontaneous long-range orders but are moderately influenced by the exchange fields from the $R$-$4f$ moment configuration \cite{Tatara2002}.
%Here we investigate the latter case, i.e., the geometrical Hall effect in the weak coupling case \cite{Tatara2002}, which is much less studied so far.
%
%Another important aspect of GHE is the relation between the mean free path and the magnetic period \cite{Onoda2004}.
%The relation between the mean free path and the magnetic period is of fundamental importance for the understanding of the geometrical Hall effect\cite{Onoda2004}.
%When the mean free path is sufficiently longer than the magnetic period, both anomalous Hall effect and GHE are appropriately captured by the momentum space picture where the anti-crossing points (e.g. Weyl nodes) are associated with the Berry curvature due to the spin-orbit coupling (Karplus-Luttinger intrinsic mechanism) and the SSC \cite{Ohgushi1999, 2003ScienceFang}.
%In contrast, as the magnetic period is longer than the mean free path, electrons hopping around the spin triad feel the emergent field $\bm{B}_{\mathrm{eff}} = h\bm{n}_{ijk}\chi_{ijk}/eS$, where $\bm{n}_{ijk}$ is the normal vector to the spin triad plane and $S$ is the area of spin triad \cite{Nagaosa2010,Nagaosa2013} (see the top panel of Fig. 1(b)).
%A representative example is the skyrmion lattice observed in chiral-lattice magnets where the size of skyrmion with quantized SSC is usually large, e.g. 10nm-100nm \cite{Ye1999,Kanazawai2011}.
\\
\ \ \ Previously, GHE in the weak-coupling regime was studied for the \tcdmo\ polycrystals \cite{Ueda2012}.
%, in which the chemical doping enables the hopping motion of conduction electrons, coupled with the local $R$-$4f$ moments.
%The conduction electrons are directly affected by the exchange field from $R$-$4f$ moments or the $f$-$d$ Kondo coupling.
It is demonstrated that the Hall response systematically changes as a function of the ``density" of SSC in the real space tuned by the doping level of magnetic Tb moments, in accord with the theoretical prediction \cite{Tatara2002}.
%, and thereby suggests the presence of the GHE even in such a weak coupling case \cite{Ueda2012}
However, the exchange-field induced GHE should be dominated by the $R$-$4f$ moment configuration; for instance, SSC is expected to change its sign and magnitude between 2-in 2-out and 3-in 1-out configuration, and further, diminish as the $R$-$4f$ moments are forced to align in a collinear manner.
Thus, to obtain a deeper insight into GHE, the field-directional study on the single crystals is required.\\
\ \ \ In this study, we investigate the magnetotransport properties of \tcmo\ single crystals to see the role of the real-space scalar spin chirality in the Hall effect.
We successfully synthesize high-quality samples by using the state-of-the-art floating zone furnace equipped with high-power lasers and measure the transport and magnetization at high magnetic fields up to 31 T, which allows us to access a wide range of Tb magnetic states from non-coplanar spin textures to fully spin-aligned state.
%In this Letter, we investigate the magnetotransport properties of the diffusive metal \tcmo to reveal the Tb 4f-moment configuration dependence of the geometrical Hall effect (GHE) in the weak $f$-$d$ coupling regime.
%We could synthesize Ca-doped \tmo single crystals and measure the transport properties and magnetization by using the high magnetic field up to 31 T, which allows us to observe the crossover from non-coplanar Tb spin textures to fully spin-aligned state.
We observe the large anisotropy of geometrical Hall effect between [100] and [111] field directions in the intermediate field range, which gradually diminishes at high fields.
It can be reasonably explained in terms of the magnitude and sign change of the scalar spin chirality of localized Tb moments that impose the {\it f-d} exchange field on the conductive Mo sublattice.
%We observe the large anisotropy of GHE in the intermediate field range, which can be assigned to be induced by the exchange-field from the respective Tb magnetic configurations with varying magnitude and sign of scalar spin chirality.
%}
\\\\
{\large{\bf Results}}\\ 
{\bf Band filling control.}
Figure 1(c) shows the temperature dependence of resistivity for several compositions. The resistivity of $x = 0$ rapidly increases as the temperature decreases. %, as reported in a previous study [? ]. 
Above $x=0.08$, the resistivity significantly decreases down to the order of $10^{-3} \ \mathrm{\Omega cm}$. The weak temperature dependence for higher doping reminds us of the high pressure effect on \nmo, which yields the anomalously diffusive metallic state presumably as a result of the strong competition between an antiferromagnetic (spin glass) insulator and ferromagnetic metal states \cite{2009PRLIguchi}. 
%The inset of Fig. 1(d) shows the optical conductivity spectra at 10 K over a wide energy range for several compositions.
%The absorption above 2 eV corresponds to the charge transfer excitations from O 2p to Mo 4d bands, and the peak at around 0.5 eV is the Mott- Hubbard gap transition. 
%As x increases, the spectral weight shifts to the lower energy across $\hbar\omega \sim 1.8$ eV (isosbestic point), which is much larger than that of the bandwidth-controlled metal-insulator transitions ($\hbar\omega \sim 0.5$ eV) \cite{Kezsmarki2006}. 
%This indicates that the \UTF{FB01}lling control modulates the band structure in the large energy scale across the insulator to metal transition. 
%
The optical conductivity spectra at 10 K for several $x$ is displayed in Fig. 1(d).
The optical conductivity for $x = 0$ gradually decreases below 0.2 eV, forming a clear charge gap of $\sim$0.05 eV.
The observed magnitude of the charge gap is slightly different from the previous one \cite{Kezsmarki2006}, possibly due to the oxygen non-stoichiometry depending on the crystal-growth condition. 
With increasing $x$, the optical conductivity gradually increases below $\omega_{\mathrm{C}}\sim $1.5 eV (see the inset of Figure 1(d)) accompanied by the closing of the charge gap, in good accordance with the dc conductivity. 
The absence of the clear Drude peak confirms the diffusive metal state at $x=0.08$ and 0.14. 
It is to be noted that the sharp peaks below 0.1 eV are infrared-active phonon modes allowed for cubic pyrochlore lattice \cite{1983JRSVandenborre,Taniguchi2004,Ueda2019}, indicative of a good crystal quality even for the doped metallic samples.

%\setmainfont{Arial}
\begin{figure}
\begin{center}
\includegraphics[width=3.375in,keepaspectratio=true]{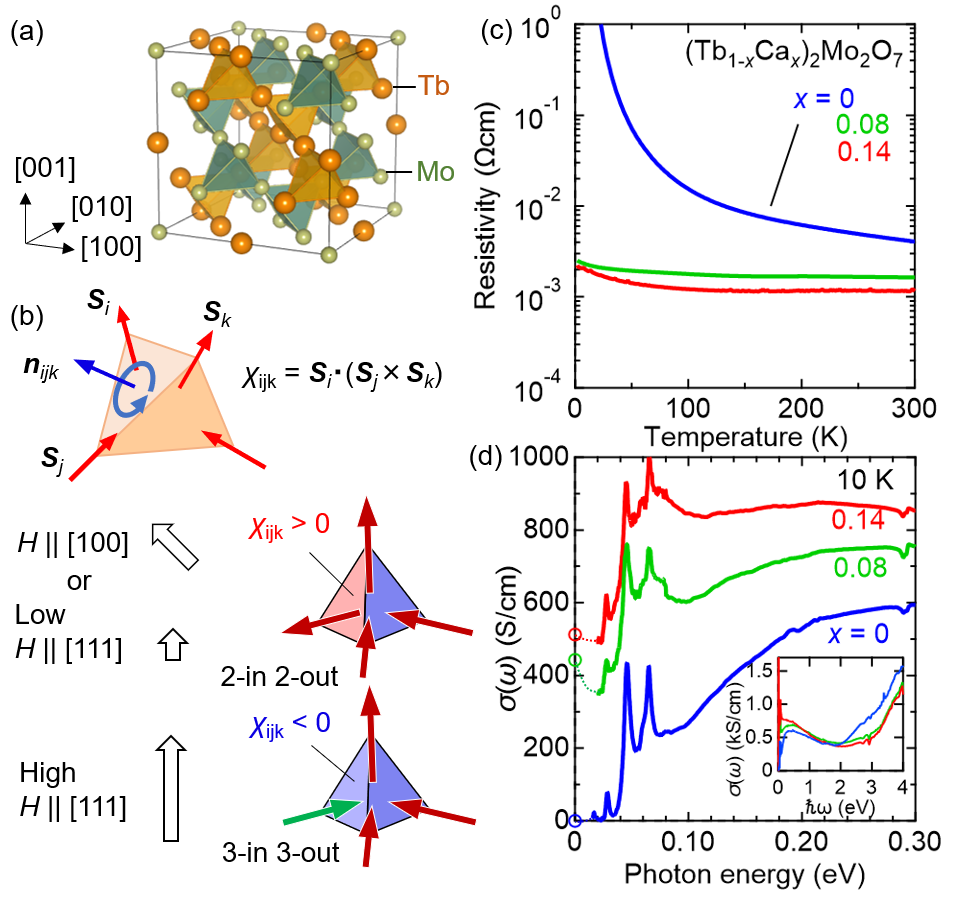}
\caption{%(colour online). 
{\bf Crystal structures and basic property in Ca-doped \tmo.}
(a) Crystal structure of pyrochlore type \tmo.
(b) Definition of the scalar spin chirality $\chi_{ijk}=\bm{S}_{i} \cdot (\bm{S}_{j}\times \bm{S}_{k})$ in a tetrahedron with Tb spins $\bm{S}_{i}$, $\bm{S}_{j}$, and $\bm{S}_{k}$ on its vertices.
The middle (bottom) figure shows $\chi_{ijk}$ in each triangle plane of a tetrahedron for 2-in 2-out (3-in 1-out) state. 
The colour of the plane corresponds to the sign of $\chi_{ijk}$ when we take $\bm{n}_{ijk} \cdot \bm{H} >0$.
In these figures, $H\parallel[111]$ case is assumed.
(c) Temperature dependence of resistivity for \tcmo.
(d) Optical conductivity spectra below 0.3 eV at 10 K for \tcmo. The inset shows the optical conductivity spectra up to 4 eV.
}
\end{center}
\end{figure}
\leavevmode\\
%\setmainfont{Times New Roman}
%Fig. 1(e) shows the phase diagram in pyrochlore type \tcmo. 
%Spin glass transition occurs at 25 K at $x = 0$, and is systematically suppressed by Ca doping (see the magnetization curve in Supplementary Material). 
%As shown in Fig. 1(e), a metallic state without a long-range magnetic order is realized by Ca doping, consistent with the previous study on the polycrystals \cite{Iguchi2011}.
%
{\bf Anisotropic magnetotransport properties.}
In the following, we focus on the magnetotransport properties of the barely metallic $x = 0.14$ crystal. 
Figure 2(b) shows the magnetic field dependence of resistivity at several temperatures for $H\parallel[111]$ and [100].
The resistivity is almost independent of magnetic field at 100 K.
Below 20 K, the resistivity gradually decreases with increasing field, likely due to the suppression of spin glass or antiferromagnetic spin fluctuation.
Figures 2(c) and (d) show the Hall resistivity $\rho_{yx}$ for $H\parallel[100]$ and $H\parallel[111]$, respectively.
$\rho_{yx}$ at 100 K is nearly proportional to the magnetic field with little difference between for $H\parallel[100]$ and $H\parallel[111]$.
However, the anisotropy becomes evident at low temperatures. At 2 K, as the field increases, $\rho_{yx}$ for $H\parallel[100]$ abruptly increases, reaches a maximum value of $\sim 9 \ \mathrm{\mu\Omega}$cm at around 8 T, and then slightly decreases. On the other hand, $\rho_{yx}$ for $H\parallel[111]$ shows a hump at around 2 T and gradually increases as the field is further increased.
As can be seen in Fig. 2(e), the anisotropy ratio of $\rho_{yx}$ between $H\parallel[100]$ and $H\parallel[111]$ at 2 K exceeds 2 in the intermediate field region, and gradually decreases as the field increases.
Figure 2(f) shows the magnetization curves for $H\parallel[100]$ and $H\parallel[111]$.
At 100 K, the magnetization shows H-linear dependence and no anisotropy is ovserved as in $\rho_{yx}$.
With lowering temperature, the magnetization for $H\parallel[100]$ becomes larger than that for $H\parallel[111]$.
Such an anisotropy can be attributed to the different magnetic configuration of Tb moments.
Since each Tb moment shows single-ion anisotropy along the local $\langle111\rangle$ orientation, the sufficiently large, but not too large, magnetic field along [100] ([111]) favors 2-in 2-out (3-in 1-out) state (see Fig. 1(b)).
In fact, the expected value of 2-in 2-out state is larger than that of 3-in 1-out state by $\sim 1.2\ \mu_{\mathrm{B}}$/f.u..
Such a difference of Tb magnetic configuration can give rise to the salient anisotropy of Hall effect.

\begin{figure}
\begin{center}
\includegraphics[width=3.5in,keepaspectratio=true]{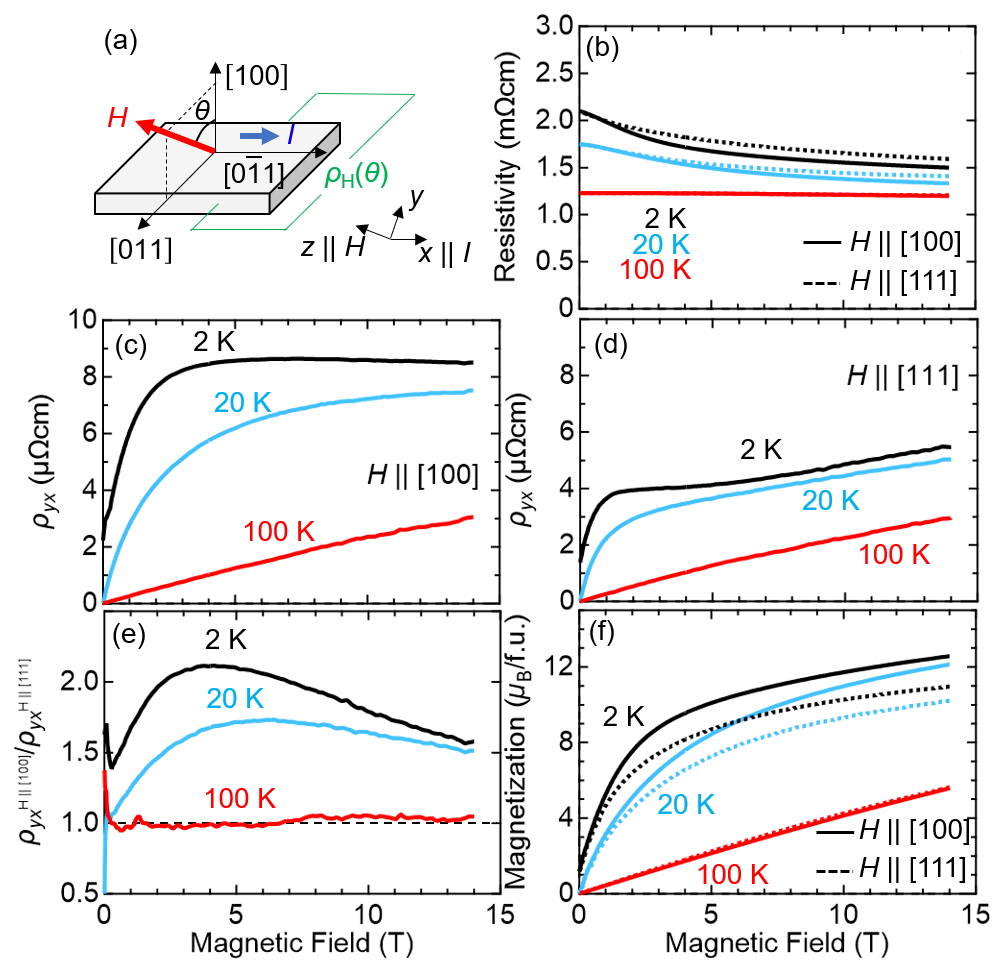}
\caption{%(colour online).
{\bf Magnetotransport properties in the diffusive metal $\bm{x=0.14}$.}
 (a) Measurement setup for magnetic transport properties.
Magnetic field dependence of (b) resistivity for both $H\parallel[100]$ and $H\parallel[111]$, (c) Hall resistivity for $H\parallel[100]$, (d) Hall resistivity for $H\parallel[111]$, (e) anisotropic ratio of Hall resistivity between $H\parallel[100]$ and $H\parallel[111]$, and (f) magnetization for both $H\parallel[100]$ and $H\parallel[111]$ at several temperatures.
%(c),(d) Magnetic field dependence of Hall resistivity. (e) Magnetic field dependence of the field-directional anisotropy of the Hall resistivity, the value for H\parallel[100] divided by the one for H\parallel [111]. (f) Magnetic field dependence of magnetization.
}
\end{center}
\end{figure}

\begin{figure}
\begin{center}
\includegraphics[width=3.5in,keepaspectratio=true]{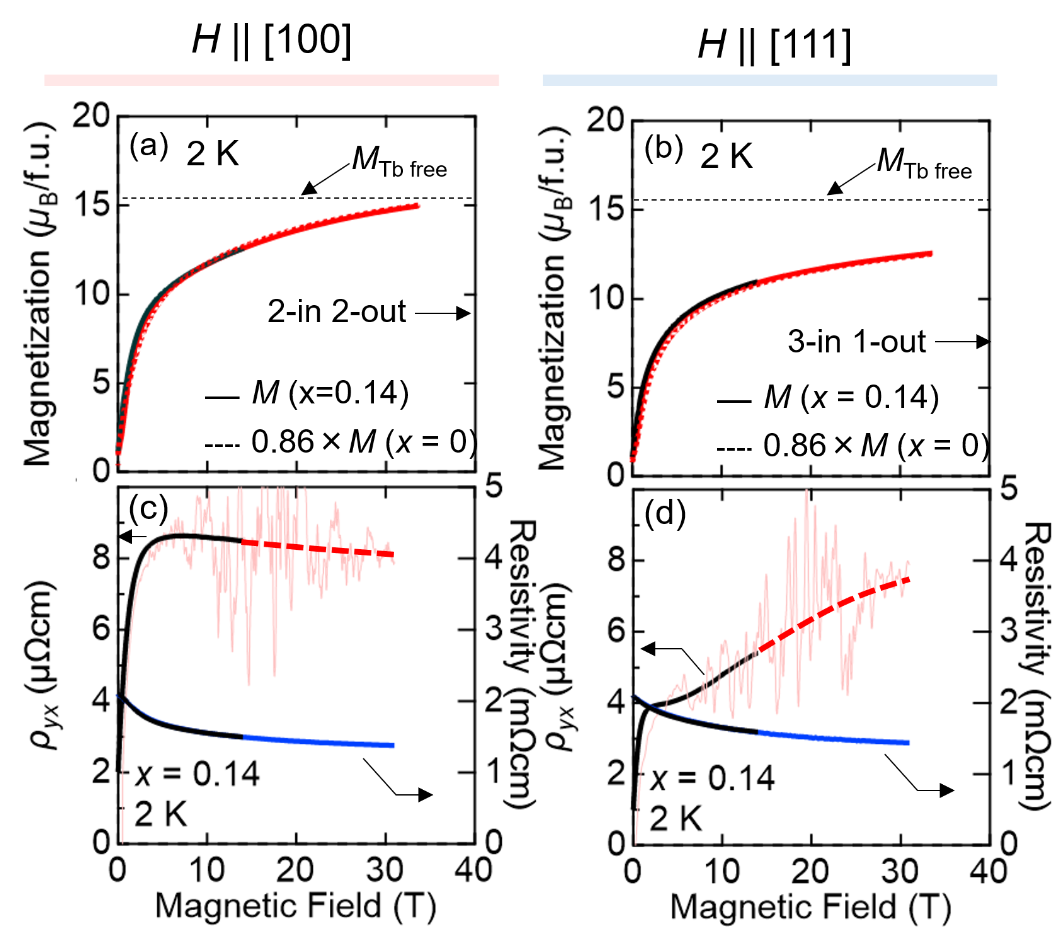}
\caption{%(colour online). 
{\bf High magnetic field measurement}
Magnetic field dependence of magnetization up to 34 T for (a) $H\parallel[100]$ and (b) $H\parallel[111]$.
Magnetic field dependence of Hall resistivity as well as resistivity for (c) $H\parallel[100]$ and (d) $H\parallel[111]$.
The black curves indicate the results of DC field measurements up to 14 T. The red and blue curves show the results of high field measurements using pulsed fields up to 31 T.
}
\end{center}
\end{figure}

\leavevmode\\
{\bf High magnetic field measurement.}
In order to elucidate the origin of Hall effect, we employ the higher magnetic field measurement which allows us to fully control the Tb-moment configurations.
Figure 3(a) (3(b)) shows the magnetic field dependence of magnetization up to above 34 T at 2 K for $H\parallel[100]$ ([111]).
The magnetization measured by the pulse magnet (red solid curve) perfectly overlaps with the one measured by the dc magnet (black curve).
Remarkably, the magnetization for $H\parallel[100]$ monotonically increases, exceeds the expected value for the 2-in 2-out state ($8.9\ \mu_{\mathrm{B}}$/f.u.) at 4 T, and nearly reaches 15.5$\ \mu_{\mathrm{B}}$/f.u. at 34 T which is expected for fully-polarized Tb-$4f$ moments, likely stemming from the competition between the magnetic anisotropy and Zeeman effect.
Thus we can examine the magnetotransport properties for both collinear and non-collinear magnetic states on one sample.
We also plot the magnetization for $x=0$ in Figs. 3(a) and 3(b).
To compare with that for $x=0.14$, we renormalize it by the density of Tb moments, namely multiply it by $1-x=0.86$. Notably, the renormalized curve for $x=0$ falls onto the same curve for $x=0.14$ above 10 T, indicating that the Mo contribution to the whole magnetization is almost negligible not only for the spin-glass insulating $x=0$ crystal but for the carrier-doped metallic $x=0.14$ one, in stark contrast to the Mo-spin ferromagnetic \nmo\ \cite{Taguchi2001}.
%Thus it is likely that the Mo conduction electrons weakly couple to the local Tb magnetic moments via the Kondo coupling and acquire the SSC inducing the GHE as well as AHE.
Figures 3(c) and 3(d) show the magnetic field dependence of resistivity and Hall resistivity for $H\parallel[100]$ and $H\parallel[111]$, respectively.
The resistivity gradually decreases as the field increases up to 31 T for both field directions, as observed in the measurement in the dc magnet.
The Hall resistivity data measured by the pulse magnet seem somewhat noisy, because of the small Hall signals compared to the longitudinal resistivity.
Nevertheless, the overall field dependence and amplitude are reconciled with the dc data, and moreover the signal appears reliable again near the maximum field, since the data can be integrated during the relatively long time at high fields in the field pulse shape.
% (see Supplementary Material). 
As the field approaches to 31 T, $\rho_{yx}$ for $H\parallel[100]$ gradually decreases down to $\sim 8 \ \mu\Omega$cm, while that for $H\parallel[111]$ explicitly increases up to $\sim 7.5 \ \mathrm{\mu\Omega cm}$. Apparently, the anisotropy of $\ryx$ disappears at high magnetic fields.
\\\\
{\large{\bf Discussion}}\\
\ \ \ In general, the Hall resistivity is expressed as
\begin{align}
\rho_{yx} = R_{\rm 0}\mu_{0}H + R_{S}\rho_{xx}^{n}M_{z} + \rho_{yx}^{\rm G}.
\end{align}
where $R_{0}$ is the ordinary Hall coefficient, $R_{S}\rho_{xx}^{n}$ the anomalous Hall coefficient with the scaling factor $n$ in case of diffusive metal \cite{Miyasato2007} , and $\rho_{yx}^{\rm G}$ the geometrical contribution.
According to the previous study on Cd-doped Y${}_2$Mo${}_2$O${}_7$, the ordinary Hall effect is quite small (less than $0.3 \ \mathrm{\mu\Omega cm}$ at 14 T).
Therefore, we neglect here the ordinary contribution for simplicity. To extract the geometrical contribution, we plot $\rho_{yx}$ as a function of magnetization in Fig. 4(a).
The red and blue solid lines denote experimental data measured in dc magnetic field, red and blue circles are the data at the highest field 31 T, and red and blue dashed lines indicate the noise-smoothed connections between these two experimental data.
Since the magnetization for $H\parallel[100]$ is almost saturated at 31 T (Fig. 3(a)), we can assume that the geometrical contribution is zero at 31 T. 
The black line in Fig. 4(a) indicates the Karplus-Luttinger type anomalous Hall resistivity, $\rho_{yx}^{\mathrm{KL}} = R_{S}\rho_{xx}^{n}M_{z}$, where $n = 0.4$ in the carrier hopping regime, as confirmed in Supplementary Figure 2\cite{Miyasato2007}.
As can be seen, $\rho_{yx}$ for $H\parallel[100]$ is larger than $\rho_{yx}^{\mathrm{KL}}$ in the whole magnetization regime. Remarkably, for $H\parallel[111]$, $\rho_{yx}$ shows non-monotonic magnetization dependence as opposed to $\rho_{yx}^{\mathrm{KL}}$. Especially, $\rho_{yx}$ crosses $\rho_{yx}^{\mathrm{KL}}$ at $M \sim 6 \ \mu_{\mathrm{B}}$/f.u., and eventually merges to $\rho_{yx}^{\mathrm{KL}}$ at the largest magnetization. 

\begin{figure}
\begin{center}
\includegraphics[width=3.5in,keepaspectratio=true]{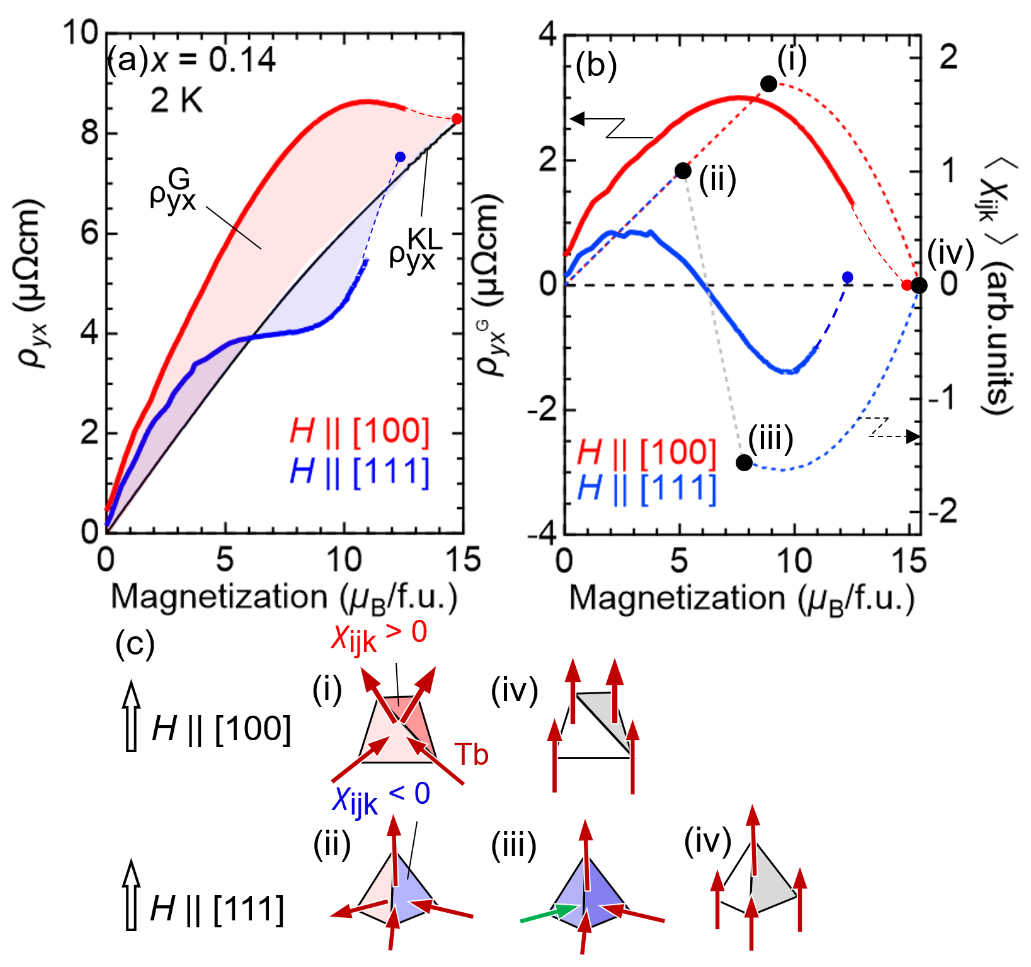}
\caption{%(colour online).
{\bf Geometrical Hall effect and Tb magnetic configuration.}
 (a) Magnetization dependence of the Hall resistivity. The solid lines denote the data obtained with the DC field, and the circles are the values at the maximum pulse field. The dotted lines are the smoothed data taken by the pulse field measurement.
(b) Magnetization dependence of the geometrical Hall effect and simulated spin chirality. 
Red and blue dashed lines represent the simulated scalar spin chirality of the single Tb tetrahedron based on the magnetization curves (see Supplementary Note 3). 
(c) Possible magnetic structure and scalar spin chirality at each point in the panel (b).
}
\end{center}
\end{figure}

Figure 4(b) displays the extracted geometrical contribution, $\rho_{yx}^{\rm G} = \rho_{yx} - \rho_{yx}^{\mathrm{KL}}$ , as a function of magnetization. 
As the magnetization increases, $\rho_{yx}^{G}$ for $H\parallel[100]$ gradually increases, reaches maximum at $\sim 8 \ \mu_{\mathrm{B}}$/f.u., and eventually decreases towards zero at the full moment of $15.5 \ \mu_{\mathrm{B}}$/f.u.
We speculate that the envelope-shaped magnetization dependence, which is also observed in several materials \cite{Ye1999, 2021PNASKolincio}, reflects the modulation of Tb magnetic states.
Both Tb and Mo magnetic moments are disordered at zero magnetic field.
As the field increases, the large Tb moments are getting aligned due to the gain of Zeeman energy under the influence of the [111] Ising anisotropy, and eventually form 2-in 2-out like configuration at $M \sim 8.9 \ \mu_{\mathrm{B}}$/f.u. (see panel (i) in Fig. 4(c)).
With further increasing field, the Zeeman energy gradually overcomes the magnetic anisotropy energy, and finally the Tb moments are fully aligned collinearly, as shown in panel (iv) in Fig. 4(c).
Here we simulate the average of SSC $\langle \chi_{ijk}\rangle$ in a single tetrahedron having four triangle planes, as shown in Fig. 1(b) (see Supplementary Note 3 for more detail).
Assuming that all Tb moments simply approach to the field direction with increasing field, we can calculate the angle between Tb moments and the external field from the magnetization value, and hence obtain $\langle \chi_{ijk}\rangle$.
Starting from the perfect 2-in 2-out state (panel (i)), $\langle \chi_{ijk}\rangle$ monotonically decreases towards zero (panel (iv)) as the magnetization is increased.
This is intuitively understandable since the solid angle subtended by Tb moments becomes smaller as the applied field is increased.\\
\ \ \ On the other hand, $\rho_{yx}^{\rm G}$ for $H\parallel[111]$ shows unique magnetization dependence.
It is somewhat similar to that for $H\parallel[100]$ below 2 $\ \mu_{\mathrm{B}}$/f.u., but $\rho_{yx}^{\rm G}$ decreases towards the negative value as the magnetization is further increased.
It takes minimum (negative maximum) at $\sim 9\ \mu_{\mathrm{B}}$/f.u. and then approaches zero at the large magnetization value. To understand this behavior, we consider the magnetic configuration and $\langle \chi_{ijk} \rangle$.
According to the neutron diffraction experiments \cite{Ehlers2010}, Tb-Tb interaction in \tmo\ is ferromagnetic, and hence favors the 2-in 2-out magnetic structure at weak magnetic fields, similar to canonical spin ice systems Dy${}_2$Ti${}_2$O${}_7$ and Ho${}_2$Ti${}_2$O${}_7$ \cite{1997PRLHarris,Sakakibara2003}, and spin-ice like ordered semimetal Pr${}_2$Ir${}_2$O${}_7$ \cite{2015PRBMacLaughlin,2022PRBUeda}.
At the intermediate field applied along [111], the apical spins, whose easy axes are along the field direction, are fixed while other three spins obey the ice rule, forming the so-called kagom\'{e} ice state (see panel (ii) in Fig. 4(c)).
Moreover, $R$ moments with the strong Ising character undergo the liquid-gas type magnetic transition from the kagom\'{e} ice state to 3-in 1-out state (panel (iii) in Fig. 4(c)) at higher fields.
We speculate that the crossover between these magnetic states occurs in the present system as well, leading to the remarkable sign change of $\langle \chi_{ijk}\rangle$ and hence of $\rho_{yx}^{\rm G}$, as shown in Fig. 4(b). 
In fact, the sign of $\langle \chi_{ijk}\rangle$ changes at $M \sim 6 \ \mu_{\mathrm{B}}$/f.u., consistent with the experimental observation of the sign change of GHE.
While the magnetization value exceeds that for the 3-in 1-out state, we anticipate that the 3-in Tb moments in the 3-in 1-out state are forcedly aligned toward the collinear state.
$\langle \chi_{ijk}\rangle$ slightly decreases and forms a broad dip centered at $\sim 10 \ \mu_{\mathrm{B}}$/f.u., and quickly increases towards zero at 15.5$\ \mu_{\mathrm{B}}$/f.u..
Despite such a simplified simulation of $\langle \chi_{ijk}\rangle$, the overall magnetization dependence of $\rho_{yx}$ can be well explained. 
The present experimental data clearly indicate that the GHE is controlled by the exchange fields from the configuration of the local Tb moments in the weak coupling regime.\\
% and thus represent the ideal platform to study the correlation between electronic transport and non-collinear magnetism.\\
%
\ \ \ In conclusion, we observe the geometrical Hall effects in the pyrochlore type \tcmo\ single crystal, whose conduction electrons are interacting with local magnetic moments in the weak coupling region, and thus represent the ideal platform to study the correlation between electronic transport and non-collinear magnetism.
The highly field-anisotropic geometrical Hall effect can be well explained by the real-space scalar spin chirality arising from the non-coplanar Tb magnetic configurations.
We clearly demonstrate that there is one-to-one correspondence between the geometrical Hall effect and scalar spin chirality, both of which vary for several magnetic states, e.g. 2-in 2-out, 3-in 1-out, and fully-aligned collinear states.
\\
% of the local exchange field arising from the real-space scalar spin chirality (SSC) of Tb tetrahedra, exemplified by the clear one-to-one correspondence of the sign changes of GHE and SSC, e.g, for 2-in 2-out vs. 3-in 1-out states. \\
%
\ \ \ 
%We are grateful to M. Hirschberger and N. Nagaosa for fruitful discussion.
%This work was supported by JSPS/MEXT Grant-in-Aid for Scientific Research(s) (grant no. 21K13871) and CREST (Grant no. JPMJCR1874) from JST.
\\
{\large{\bf Methods}}\\ 
{\bf Single crystal growth.}
%Single crystals \tcmo\ were synthesized by using the floating zone furnace with lasers \cite{Kaneko2020}.
%Firstly, Tb${}_2$O${}_3$, CaCO${}_3$, and MoO${}_2$ were mixed in stoichiometric ratio, pressed into pellet, and baked at 1000${}^{\circ }$C for 24 hours under argon atmosphere. After cooling to the room temperature, it was well grounded, and pressed into rods, and baked again at 1400${}^{\circ }$C for 24 hours under argon atmosphere.
%method
%Single crystals were synthesized at $\sim 1880 {}^{\circ}$C under an argon atmosphere of 0.99 MPa in the laser floating zone furnace \cite{Kaneko2020}.
%All of them were well characterized by x-ray powder diffraction and energy dispersive x-ray to check the lattice structure and composition.
%The crystal orientation was determined by back-Laue pattern. For $x > 0.20$, polycrystalline samples were also synthesized by the high-pressure furnace. A part of the crushed sintered rod was packed in a platinum capsule, and heated at 1200${}^{\circ}C$ and 4.5 GPa for 20 minutes \cite{Ueda2012}.
Single crystals \tcmo\ were synthesized by using the state-of-the-art ﬂoating zone furnace equipped with high-power lasers \cite{Kaneko2020}. Firstly, Tb${}_2$O${}_3$, CaCO${}_3$, and MoO${}_2$ were mixed in stoichiometric ratio, pressed into pellet, and baked at 1000 ${}^{\circ}\mathrm{C}$ for 12 hours under argon atmosphere. After cooling to the room temperature, it was well grounded, and pressed into rods, and baked again at 1400 ${}^{\circ}\mathrm{C}$ for 18 hours under argon atmosphere. The sintered rod was heated at $\sim$1880 ${}^{\circ}\mathrm{C}$ under an argon atmosphere of 0.99 MPa in the laser ﬂoating zone furnace. The single crystals were well characterized by x-ray powder diffraction and energy dispersive x-ray to check the lattice structure and composition. The crystal orientation was determined by back-Laue pattern. 
\\\\
{\bf Transport and magnetization measurements.}
The transport properties and magnetization up to 14 T were measured by using dc magnets in Physical Property Measurement System (PPMS), Quantum Design.
Higher-field measurements of magnetotransport and magnetization were performed by using non-destructive pulse magnets energized by capacitor banks and a flywheel DC generator installed at International MegaGauss Science Laboratory of Institute for Solid State Physics (ISSP), University of Tokyo.
Resistivity and Hall conductivity were measured by a five-probe method.
Transport measurement setup is shown in Fig. 2(a). The electric current flows along $[0\overline{1}1]$ crystalline direction. The magnetic field was rotated around the current direction, so that the transport measurements for different orientations can be performed on the same sample. $\theta = 0$, which is perpendicular to the sample plane, is along [100] crystalline direction, while $\theta = \pm 54.7^{\circ}$ is along the [111] and its equivalent direction.
We obtained the Hall resistivity $\rho_{yx}$ by normalizing the measured Hall signal $\rho_{H}$ for $\theta = \pm 54.7^{\circ}$ divided by $\cos\theta$.
We confirm that the deduced $\rho_{yx}$ shows almost the same behavior as another sample with the [111] plane measured under the normal [111] field (see Supplementary Figure 1).
\\\\
{\bf Optical measurement.}
Reflectivity spectra for vertical incidence light were measured in the range of 0.008 to 5 eV.
We used Fourier transform spectrometers for infrared region and grating-type ones for visible to ultraviolet region. 
Room-temperature spectra measurement up to 40 eV were performed by using synchrotron radiation at UV-SOR, Institute for Molecular Science. 
Optical conductivity spectra were obtained by Kramers-Kronig analysis.
As for a lower energy ($<$0.008 eV) region, we used constant reflectivity extrapolation for $x=0$ and Hagen-Rubens relation for $x=0.08,0.14$.
As for a higher energy ($>$40 eV) region, we assumed $R \propto E^{-4}$ where $R$ is the reflectivity and
$E$ is photon energy.
\\\\
{\large{\bf Supplementary Material}}\\ 
{\bf S1. Hall resistivity measurement.}
%In the main text, it is shown that anisotropy appears in the magnetotransport properties in the [100] and [111] directions.
The magnetic anisotropy of Hall effect was investigated by rotating the sample A against the applied magnetic field, so that we can rule out the possibility of sample dependence.
Fig. S1(a) shows the geometry of the sample A.
It was polished into the plate-like shape whose vertical direction is along [100] crystalline axis. 
The electric current was flowed along [0$\bar{1}$1] axis, about which the magnetic field was rotated.
The transverse resistivity $\rho _{\rm H}$ was measured along [011] axis which is perpendicular to both [100] and [0$\bar{1}$1] axes.
%The reason why we did not compare two samples, one is like FIG. S1(a) and other is like FIG. S1(b), is to avoid sample dependence.
%However, it should be noted that in the [111] direction, unlike the [100] direction, the voltage reading is not perpendicular to the magnetic field and current.

%Now we will discuss the effect of rotating magnetic field.
%We use the setup like FIG. S1(a) or FIG. S1(b).
Let us take the magnetic field direction as the $z$-axis and the current direction as the $x$-axis, as depicted in Fig. S1(a).
When the field is tilted by $\theta $ off [100] direction, Hall resistivity $\rho_{\rm H}$ is expressed as
\begin{align}
\label{eq.1}
\rho_{\mathrm{H}}(\theta) = \rho_{yx}(\theta)\cos\theta - \rho_{zx}(\theta)\sin\theta.
\end{align}
%$w$ is the width of the sample and j_{x} is the current density along x-axis.
The first term is the usual Hall effect perpendicular to both magnetic field and current, which should be extracted from the measured value $\rho_{\rm H}$.
%Second term is a kind of planar Hall effect.
%To extract only first term, we did following analysis.
Considering the symmetry of the setup of Fig. S1(a), we obtain $\rho_{yx}(\theta)=\rho_{yx}(-\theta)$ and $\rho_{zx}(\theta)= \rho_{zx}(-\theta)$, while $\sin \theta $ is odd.
Therefore, we can extract $\rho_{yx}(\theta)$ by averaging $\rho _{\rm H}(+\theta)$ and $\rho _{\rm H}(-\theta)$ and then dividing by $\cos\theta$.
Fig. S1(c) displays the magnetization dependence of $\rho _{yx}$ for both [100] and [111] field directions, in which one can clearly see the anisotropy, as discussed in the main text.

To double-check $\rho _{yx}$ for $H/[111]$, we prepare another sample B, whose out-of-plane axis is along [111] crystalline direction while the current direction is the same as that in sample A ([0$\bar{1}$1]).
%So we can not extract only $\rho_{yx}$ from $V_{\mathrm{H}}(\theta)$ when $\theta \neq 0$, which means we can not obtain [100] direction under the setup of FIG. S1(b).
%This is the reason why we used FIG. S1(a).
In this geometry, we can obtain $\rho _{yx}=\rho _{\rm H}$ for $H//[111]$ without any calculation, as plotted in Fig. S1(c).
Although the chemical composition of sample B ($x=0.134$) is slightly different from that of sample A ($x=0.140$), $\rho _{yx}$ of sample B is remarkably similar to that of sample A for $H//[111]$.
%But by using FIG. S1(b) setup, we can obtain H // [111] Hall resistivity without processing the data.
%If the H // [111] Hall resistivity measured in FIG. S1(a) and FIG. S1(b) show same behavior, it means our analysis correctly extracts H//[111] Hall resistivity under FIG. S1(a) setup.
Thus, the anisotropy discussed in the main text is ubiquitous for the present system.

\begin{figure}
\begin{center}
\includegraphics[width=3.375in,keepaspectratio=true]{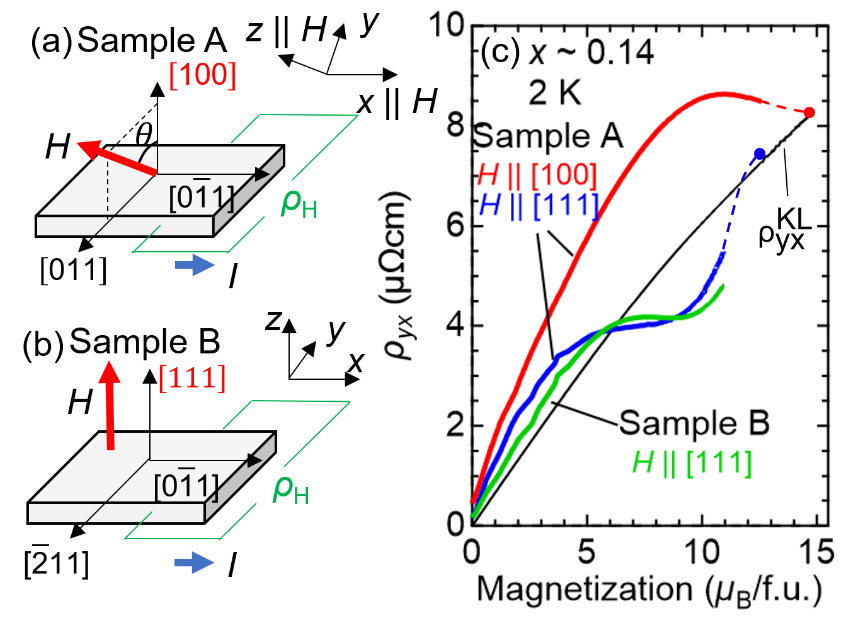}
\caption{
%(color online). (a),(b) Measurement setups. (a)/(b) is cut out to surface be perpenducular to the [100]/[111]  direction. 
%(c),(d) Magnetization dependence of Hall resistivity in the (a)/(b) setup.
%Both shows same magnetization dependence.
%(color online). 
Measurement setups for (a) sample A which is mainly used for the anisotropic measurement and (b) sample B for comparison. Each out-of-plane axis is along the [100] and [111] crystalline orientation, respectively.
(c) Magnetization dependence of Hall resistivity for (a) and (b) setups.
The red and blue curves are measured for sample A, and the green curve is measured for sample B. The black curve is Karplus-Luttinger term of Hall resistivity. For more detail, see the main text.
%Both shows same magnetization dependence.
}
\end{center}
\end{figure}

\leavevmode\\
{\bf S2. Scaling Law for Hall Conductivity.}
In the hopping region, the anomalous Hall conductivity $\sigma_{xy}$ is known to follows the empirical scaling relation $\sigma_{xy}\propto \sigma_{xx}^{1.6}$ where $\sigma_{xx}$ is longitudinal conductivity \cite{Miyasato2007}.
Fig. S2 shows $\sigma _{xy}$ normalized by magnetization as a function of $\sigma _{xx}$ for several different samples at 2 K and 14 T, at which the anomalous Hall contribution is dominant.
%The Hall conductivity includes all ordinally, anomalous, and geometrical Hall effects, but as mentioned in the previous section, the normal Hall effect can be neglected.
%In addition, the value of the geometrical Hall effect is made as small as possible by applying a magnetic field of 14 T to make the magnetic structure collinear.
%Therefore, the majority of the Hall effect is comming from the term which is proportional to the magnetization.
%Since the magnitude of the magnetization is different when the magnetic field is applied in the [100] or [111] direction, the Hall conductivity is normalized by the magnetization.
One can see that $\sigma_{xy} \propto \sigma_{xx}^{1.71}$ holds for a wide range of $\sigma _{xx}$ from 10$^{2}$ S/cm to 10$^{3}$ S/cm.
Therefore, we use $n=0.4$ in eq. (1) in the main text to estimate the Karplus-Luttinger term $\rho _{yx}^{\rm KL}$.
%The deviation from the empirical $\sigma_{xy} \propto \sigma_{xx}^{1.6}$ is assumed to be due to subtle included geometrical Hall effects.
However, the $H$-dependence of $\sigma_{xx}$ is not large, and even if there were a slight discrepancy in $n$, the conclusion shown in Fig.4 of the main text would be hardly affected.

\begin{figure}
\begin{center}
\includegraphics[width=3.375in,keepaspectratio=true]{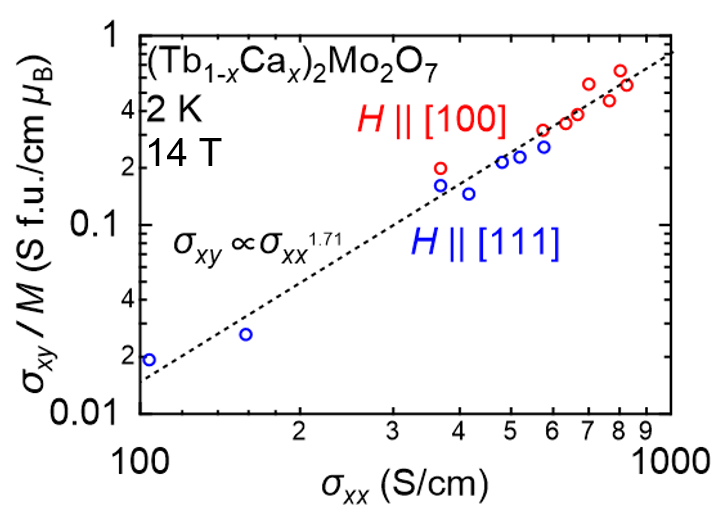}
\caption{%(color online). 
Scaling plot of $\sigma_{xy}$ versus $\sigma_{xx}$ in the logarithmic scale. The red (blue) marks denote the data for $H//[100]$ ($H//[111]$). The dashed line indicate the scaling relation $\sigma_{xy} \propto \sigma{xx}^{1.71}$.
% It shows $\sigma_{xy} \propto \sigma{xx}^{1.71}$ scaling, which is near to the $\sigma_{xy} \propto \sigma_{xx}^{1.6}$ known to hold for the hopping regime.
}
\end{center}
\end{figure}
\ \ \ 
\leavevmode\\
{\bf S3. Calculation of the scalar spin chirality.}
Since the Mo magnetization in \tcmo is negligible, as demonstrated in Fig. 2 of the main text, we can speculate the magnetic structure of the Tb tetrahedra from the magnetization and hence calculate the scalar spin chirality $\chi_{ijk}$.

%Now we calculate the magnetic structure by magnetization data.
%As mentioned in the main text, 
First we consider the case for $H//[100]$.
%Below the 2I2O magnetization $M_{\mathrm{2I2O}}=10.4(1-x)$[$\mu_{B}$/f.u], we assume perfect ising anistropy of Tb moments.
%This is a bit strong assumption, but the Ising anisotropy of Tb moment in pyrochlore lattice is reported \cite{}.
At zero magnetic field, Tb magnetic moments are disordered but host 2-in 2-out (2I2O) like habit because of the ferromagnetic interaction between nearest-neighbor Tb moments.
%2I2O Tb tetrahedron have randomly oriented magnetization at zero magnetic field, thus total magnetization is zero.
As the magnetic field increases, the number of 2I2O Tb tetrahedra, whose net magnetizations are parallel to the field direction,  increases and hence the magnetization reaches the expected value of $M_{\mathrm{2I2O}}=2(1-x)\times (\frac{4\mu _{\rm Tb}}{\sqrt{3}})/4=10.4(1-x)$ [$\mu_{B}$/f.u] where $\mu _{\rm Tb} = 9$ [$\mu_{B}$/Tb] %\cite{1999PRLGardner} 
(state (i) in Fig. 4(c) of the main text).
%And this 2I2O tetrahedra density $x_{\mathrm{2I2O}}$ is given by $x_{\mathrm{2I2O}} = M/M_{\mathrm{2I2O}}.$
% increases in the field direction up to $M_{\mathrm{2I2O}}=10.4(1-x)$[$\mu_{B}$/f.u].
%When a magnetic field is applied in the [100] direction, the magnetization increases as the number of Tb tetrahedra with 2I2O magnetic structure increases in the field direction up to $M_{\mathrm{2I2O}}=10.4(1-x)$[$\mu_{B}$/f.u].
With increasing field above 2 T, we speculate that the Zeeman energy overcomes the single-ion anisotropy and hence the Tb moments gradually approach the field direction.
%When the magnetization is above $M_{\mathrm{2I2O}}$, magnetization increases due to the bending of the Tb moment in the magnetic field direction \cite{Yin2013}, as shown in FIG. S4(b).
We define $\theta$ as the angle between the Tb moments and the magnetic field direction (Fig. S3(a)), so that the magnetization is given as $M_{\mathrm{100}}  = 2(1-x)\mu_{\mathrm{Tb}}\cos\theta $.
%The magnetization per one Tb ion in \tcmo\ is given as
%\begin{align}
%\label{eq.3}
%M_{\mathrm{H100}} & = (1-x)/2 \times 4\mu_{\mathrm{Tb}}\cos\theta
%\end{align}
%Note that $(1-x)/2$ is multiplied because to replace the magnetization in \tcmo\.
In this way, we estimate the tilt angle of the Tb moment from the measured magnetization to calculate $\chi_{ijk}$ as discussed later.

Magnetization process for $H//[111]$ is more complex. 
Similar to spin-ice materials, the magnetization for [111] field direction increases with increasing field towards $M_{\mathrm{KI}}=2(1-x)\times (\mu _{\rm Tb}+\frac{\mu _{\rm Tb}}{3})/4=6(1-x)$ [$\mu_{B}$/f.u], at which the kagome ice state is realized (the state (ii) in Fig. 4(c) in the main text).
In other words, the 2-in 2-out state is preserved up to $M_{\mathrm{KI}}$, resulting in the nearly the same curve as the case for $H//[100]$.
%1. Below the spin ice magnetization $M_{\mathrm{ice}}=6(1-x)$[$\mu_{B}$/f.u]: 2I2O Tb tetrahedron with positive magnetization in the direction of magnetic field increase. 
%2IIO tetrahedra density $x_{\mathrm{ice}}$ is given by $x_{\mathrm{ice}} = M/M_{\mathrm{ice}}.$
Above $M_{\mathrm{KI}}$, the crossover occurs from (ii) 2-in 2-out state to (iii) 3-in 1-out state, at which the magnetization becomes $M_{\mathrm{3I1O}}=2(1-x)\times (\mu _{\rm Tb}+\frac{\mu _{\rm Tb}}{3}\times 3)/4=9(1-x)$ [$\mu_{B}$/f.u].
%2. Above the spinice magnetization and Below the spin 3I1O magnetization $M_{\mathrm{3I1O}}=9(1-x)$[$\mu_{B}$/f.u]: Spin ice to 3I1O transition occurs \cite{Sakakibara2003}.
Eventually, as the field increases further, Tb moments deviate from the local $\langle 111\rangle$ axes and get aligned collinearly (the state (iv) in Fig. 4(c) of the main text).
Between (iii) and (iv) states, the magnetization is written as $M_{\mathrm{111}} = 2(1-x)\times (\mu_{\mathrm{Tb}} + 3\mu_{\mathrm{Tb}}\cos\theta)/4$ where $\theta $ is the angle between Tb moments and the field, as shown in Fig. S3(b).

Making use of the information of Tb magnetic configurations as extracted above, we calculate the spin chirality of the Tb tetrahedron averaged over four Tb sites on the vertices of one tetrahedron.
According to the theoretical calculation in the weak coupling region \cite{Tatara2002}, the emergent magnetic field $H_{\mathrm{eff}}$ acting on the Mo-conducting electrons is given as
%\begin{align}
%\label{eq.5}
%H_{\mathrm{eff}} & \propto \langle \chi_{ijk}\rangle = \sum_{i,j,k \in 1,2,3,4} \bm{S}_{i}\cdot(\bm{S}_{j}\times\bm{S}_{k})\bm{n}_{ijk}\cdot \bm{e}_{z}.
%\end{align}
\begin{flalign}
H_{\mathrm{eff}} &\propto \langle \chi_{ijk}\rangle \notag \\
{}& =  \frac{1}{6N}\sum_{i,j,k \in \mathrm{all \  sites}} \bm{S}_{i}\cdot(\bm{S}_{j}\times\bm{S}_{k})\bm{F}(r_{i},r_{j},r_{k})\cdot\bm{e}_{z} &
\end{flalign}
where
\begin{flalign}
\bm{F}(\bm{r_{i}},\bm{r_{j}},\bm{r_{k}})
{}&= (\bm{a}\times\bm{b}/ab)I^{\prime}(a)I^{\prime}(b)I(c) \notag 
\\{}&+(\bm{b}\times\bm{c}/bc)I(a)I^{\prime}(b)I^{\prime}(c) \notag 
\\{}&+(\bm{c}\times\bm{a}/ca)I^{\prime}(a)I(b)I^{\prime}(c)
\end{flalign}
$\bm{e}_{z}$ is the unit vector parallel to the field direction, $\bm{r}_{i}$, $\bm{r}_{j}$, and $\bm{r}_{k}$ are the positions of local spins at site $i$,$j$, and $k$, $\bm{a}=\bm{r}_{i}-\bm{r}_{j}$, $\bm{b}=\bm{r}_{j}-\bm{r}_{k}$, $\bm{c}=\bm{r}_{k}-\bm{r}_{i}$ are the distances between each site, $I(\bm{r})$ is the rapidly decreasing function of $\bm{r}$, and $I^{\prime}=\frac{dI(r)}{dr}$.
Thus we take into account only the nearest neighbor sites 1, 2, 3, and 4 in a single tetrahedron shown in Fig. S3(c)
\begin{align}
\label{eq.5}
\langle \chi_{ijk}\rangle 
& \propto \sum_{i,j,k \in 1,2,3,4} \bm{S}_{i}\cdot(\bm{S}_{j}\times\bm{S}_{k})\bm{n}_{ijk}\cdot \bm{e}_{z}.
\end{align}
%FIG. S4(d) shows each quantity in (\ref{eq.5}). 
where $\bm{n}_{ijk}$ is the unit vector defined by right hand screw rule when orbiting the site $i$, $j$, and $k$.
%The sum of these for all faces of the tetrahedra acts as a virtual magnetic field. 
%We take the inner product of $\bm{e}_{z}$ to extract only the magnetic field direction component.
The calculated results are plotted in Fig. 4(c) in the main text.
%Below $M_{\mathrm{2I2O}}$ (in the [100] magnetic field) or $M_{\mathrm{ice}}$ (in the [111] magnetic field), 2I2O Tb tetrahedra are present in $x_{\mathrm{2I2O}}$ or $x_{\mathrm{ice}}$ ratio in all lattices.
%It produces $\langle \chi_{ijk}\rangle = x_{\mathrm{2I2O}}\langle \chi_{ijk}^{\mathrm{2I2O}}\rangle$ or $\langle \chi_{ijk}\rangle = x_{\mathrm{ice}}\langle \chi_{ijk}^{\mathrm{ice}}\rangle$ spin chirality on average. Here, $\langle \chi_{ijk}^{\mathrm{2I2O}}\rangle$ or $\langle \chi_{ijk}^{\mathrm{ice}}\rangle$ are perfectly aligned 2I2O or spin ice state spin chirality.
We note that the spin chirality is not calculated but just a connecting straight line drawn between $M_{\mathrm{KI}}$ and $M_{\mathrm{3I1O}}$ in $H//[111]$ (the state (ii) and (iii) in Figs. 4(b) and 4(c)), because the detailed magnetic state is not trivial.
\begin{figure}
\begin{center}
\includegraphics[width=3.5in,keepaspectratio=true]{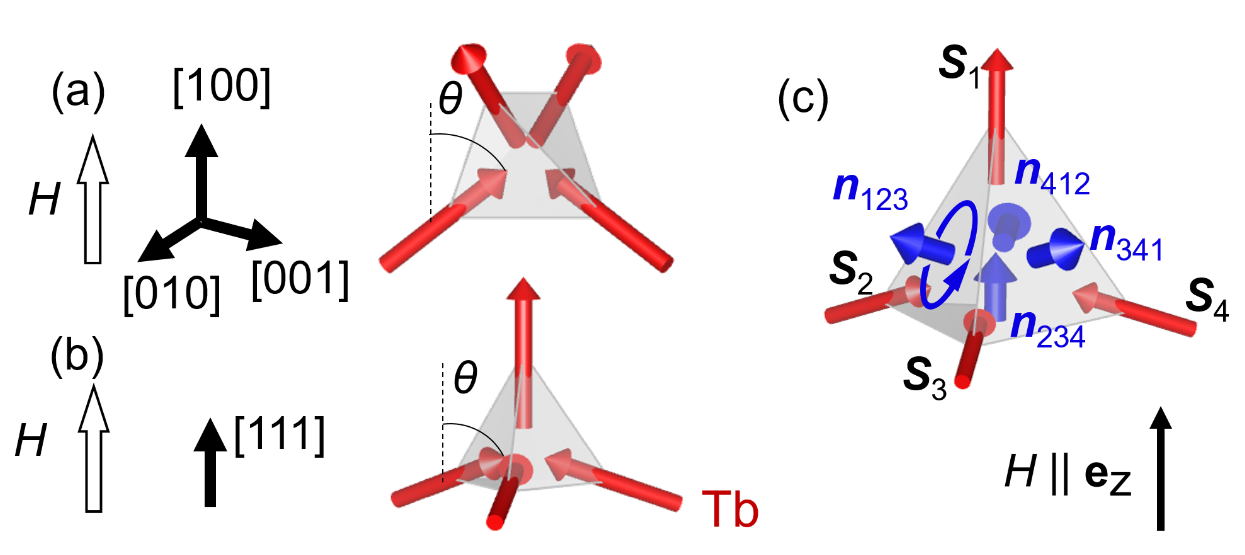}
\caption{%(color online).
%(a) Magnetization process by magnetic field in [100] direction at $M \leq M_{\mathrm{2I2O}}$. 
Tb moments modulation for (a) $H//[100]$ at $M \geq M_{\mathrm{2I2O}}$ and (b) $H//[111]$ at $M \geq M_{\mathrm{3I1O}}$. 
(c) Schematic picture of the tetrahedron for the calculation of the scalar spin chirality. Red arrows indicate the spins at each site  ($\bm{S}_{1}$, $\bm{S}_{2}$, $\bm{S}_{3}$, and $\bm{S}_{4}$), and blue arrows indicate the unit vector parallel to the emergent field ($\bm{n}_{123}$, $\bm{n}_{234}$, $\bm{n}_{341}$, and $\bm{n}_{412}$).
}
\end{center}
\end{figure}
%So we just draw a straight line between spin ice point and 3I1O point.
%Above $M_{\mathrm{2I2O}}$ (in the [100] magnetic field) or $M_{\mathrm{3I1O}}$ (in the [111] magnetic field), Tb moments are forcibly ordered (but tilted from [111] easy axis).
%We can calculated spinchirality by \ref{eq.5}.
%We will discuss the details of the calculation method of spin chirality, which we did not go into in the main text.
The calculated curves of the SSC based on the above simple assumption reasonably reproduce the experimental results of geometrical Hall components. 
 %although they are based on the very simple assumption.
%The transport properties and magnetization up to 14 T were measured by using dc magnets in Physical Property Measurement System (PPMS), Quantum Design.
%Higher-field measurements of magnetotransport and magnetization were performed by using non-destructive pulse magnets energized by capacitor banks and a flywheel DC generator installed at International MegaGauss Science Laboratory of Institute for Solid State Physics (ISSP), University of Tokyo.
%{\bf Optical conductivity measurement}
%
%\makeatletter
%\renewcommand{\@biblabel}[0]{$\ast$}
%\makeatother
\\\\
{\bf Data availability.} The data that support the findings of this study are available from the corresponding author upon reasonable request.

\leavevmode\\
{\large{\bf Acknowledgements.}}\\
\ \ We would like to thank enlightning discussion with Max Hirschberger.
This work was supported by JSPS/MEXT Grant-in-Aid for Scientific Research(s) (grant no. 21K13871) and CREST (Grant no. JPMJCR1874) from JST.
\\\\
{\large{\bf Author contributions}}\\
\ \ Y.To. conceived and guided the project. H.F. and Y.K. performed single crystal growth.
H.F. measured transport, magnetization, optical conductivity up to 14 T with help from K.U., K.K. and Y.Ta..
H.F., R.K., A.M. and M.T. performed high field measurement of transport and magnetization.
All authors discussed the results and contributed to the manuscript.
%
%{\large{\bf Additional information}}\\
%{\bf Supplementary Information} accompanies this paper at \\ http://www.nature.com/naturecommunications 
%\\\\
{%\bf Competing financial interests.} The authors declare no competing financial interests.
%\\\\
%{Materials \& Correspondence.} Indicate the author(s) to whom correspondence and material requests should be addressed.
%\\\\
{%\bf Reprints and permission information} is available online at \\http://npg.nature.com/reprintsandpermissions/
%How to cite this article: Kondo, T. et al. Quadratic Fermi node in a 3D strongly correlated semimetal. Nat. Commun. 6:10042 doi: 10.1038/ncomms10042 (2015)
\newpage
\end{document}
\bye